\documentclass[twocolumn,preprintnumbers,amsmath,amssymb,superscriptaddress,footnote]{revtex4-1}
\usepackage{graphicx} 
\usepackage[dvipdfmx]{epsfig} 
\usepackage{bm}
\usepackage{ulem}
\usepackage{amsmath,color}

\begin{document}
\title{Enlarged deformation region in neutron-rich Zr isotopes by the second intruder orbit}
\author{W. Horiuchi}
\email{whoriuchi@omu.ac.jp}
\affiliation{Department of Physics, Osaka Metropolitan University, Osaka 558-8585, Japan}
\affiliation{Nambu Yoichiro Institute of Theoretical and Experimental Physics (NITEP), Osaka Metropolitan University, Osaka 558-8585, Japan}
\affiliation{RIKEN Nishina Center, Wako 351-0198, Japan}
\affiliation{Department of Physics, Hokkaido University, Sapporo 060-0810, Japan}

\author{T. Inakura}
\affiliation{Laboratory for Zero-Carbon Energy, Institute of Innovative Research, Tokyo Institute of Technology, Tokyo 152-8550, Japan}

\author{S. Michimasa}
\affiliation{Center for Nuclear Study, The University of Tokyo, 2-1 Hirosawa, Wako, Saitama 351-0198, Japan}

\author{M. Tanaka}
\affiliation{RIKEN Nishina Center, Wako 351-0198, Japan}

\preprint{NITEP 171}

\begin{abstract}
  Nuclear deformations and density profiles
  of neutron-rich even-even Zr isotopes are investigated
  by using the Skyrme-Hartree-Fock-Bogoliubov method.
  Large quadrupole and hexadecapole deformations
  are predicted along with large enhancement
  of the total reaction cross sections
  at the neutron number $N= 60$--74.
  Strong nuclear deformation starting at $N=60$ is induced by the occupation of
  the intruder orbit with the asymptotic quantum number
  $[nn_z\Lambda]\Omega=[550]1/2$ originating
  from the spherical $0h_{11/2}$ orbit.
  The deformation region is further enlarged from $N= 72$ to 74
  owing to the occupation of the next intruder orbit with $[530]1/2$
  originating from the spherical $1f_{7/2}$ orbit.
  This characteristic nuclear deformation 
  is crucially reflected in a systematic behavior
  of the nuclear radii and the density profiles near the nuclear surface.
\end{abstract}
\maketitle

Nuclear deformation gives a significant impact on the change of the nuclear density profiles near the nuclear surface, 
resulting in the enhancement of the nuclear radii~\cite{Minomo11,Takechi12,Sumi12,Minomo12,Horiuchi12,Takechi14,Watanabe14,Horiuchi15,Horiuchi21,Choudhary21, Horiuchi22}. 
In the nuclear chart, there exist some regions where a nuclear system
exhibits unexpectedly large deformation. 
The most striking examples are
the island-of-inversion nuclei around
the neutron numbers $N=20$~\cite{Warburton90}
and $40$~\cite{Lenzi10},
where the ordering of the single-particle orbits
near the Fermi level is inverted by the intruder orbit coming
from a higher shell in the spherical limit.
The intruder orbit is highly elongated along
the quantization axis in the intrinsic frame and thus its occupation induces
a deformed nuclear shape, i.e., diffused nuclear surface
in the laboratory frame. Theoretical analyses~\cite{Choudhary21,Horiuchi22}
showed that the shell evolution in the island-of-inversion regions
results in diffusing the density profile near the nuclear surface
and increases the nuclear radii.
Here we study the development of nuclear deformation
in $_{40}$Zr isotopes, where the intruder configuration
is expected to play a crucial role throughout the isotope chain.

Zr isotopes exhibit complicated isotope dependence on nuclear deformation.
Experiments confirmed isotopes between $^{78}$Zr~\cite{Kienle01} and $^{114}$Zr~\cite{Sumikama21} and theories predicted that Zr isotope persists
to at least $^{128}$Zr~\cite{Erler12}. 
The neutron-deficient $N=Z$ nucleus $^{80}$Zr
was predicted having largely
deformed ground state~\cite{Yamagami01, Llewellyn21}.
The stable nuclei $^{90-96}$Zr and $^{98}$Zr are spherical,
and $^{100}$Zr becomes again largely deformed ground state.
The large quadrupole deformation in $^{100-106}$Zr
was confirmed experimentally by measurements of the electric-quadrupole ($E2$)
transition strengths in even-$N$ isotopes~\cite{NuDat3, Browne15}
and the $E2$ moments for odd-$N$ isotopes~\cite{Stone21}.
Furthermore, the measurements of the first $2^+$ excitation energies
in $^{108,110}$Zr strongly indicate that the large deformation persists
up to $^{110}$Zr~\cite{Sumikama11,Paul17}. 
The quadrupole deformations along the Zr isotope chain
were theoretically evaluated
by mean-field~\cite{Geng03,RodriguezGuzman10,Nomura16,Miyahara18} 
and shell model~\cite{Sieja09, Togashi16} approaches.
Some Zr isotopes are candidates
of tetrahedral nuclei~\cite{Tagami15,Zhao17} and/or shape coexistence phenomena~\cite{Kumar14,Kumar21,Abusara17,GarciaRamos19a,GarciaRamos19b}.

To extract the quadrupole deformation of unstable nuclei,
the reduced $E2$ transition strength $B(E2)$ from
the first $2^+$ state is one of the standard physical quantities
for even-even nuclei.
The $B(E2)$ value is experimentally deduced
by measuring angular distributions of the cross sections
of the Coulomb excitation or the lifetime of the first $2^+$ state.
Hadronic probes through nuclear interactions can also be used.
The proton inelastic scattering at intermediate energy is theoretically established as a probe of the quadrupole deformation~\cite{Pinkston61},  
and this method was successfully adopted for many experiments
for exotic radioactive nuclei~\cite{Yanagisa03, Takeuchi09, Aoi09, Aoi10, Michimasa14}. 

Here, we focus on the observation of
the total reaction cross section ($\sigma_R$)
as a measure of quadrupole deformation.
The $\sigma_R$ or interaction cross section measurement
has often been performed to extract the size of nuclei, especially in nuclei
far from the $\beta$ stability line because
the cross section is extremely large by $\approx 1$~barn.
Furthermore, the theoretical framework linking the cross section
to the nuclear matter radius is well-tested and established
~\cite{NTG,Varga02,Suzuki03,Horiuchi06,Horiuchi07,Ibrahim09,Takechi09,Horiuchi10,Horiuchi12,Horiuchi15,Horiuchi16,Horiuchi17,Nagahisa18}. Therefore, $\sigma_R$ is a promising observable
for detecting anomalies of the nuclear structure appearing 
along isotopic chains~\cite{Tanihata85,Fukuda91,Suzuki95,Fukuda99,Suzuki99,Ozawa00,Ozawa01,Fang04,Yamaguchi04,Yamaguchi08,Tanaka10a,Tanaka10b,Kanungo11,Yamaguchi11,Takechi12,Moriguchi13,Moriguchi14,Takechi14,Togano16,Bagchi20,Tanaka20}.
The present study is strongly linked to
the achievements in the development of nuclear deformation
in the island-of-inversion nuclei,
Ne and Mg isotopes~\cite{Takechi12,Takechi14},
where the strong quadrupole deformation starting at around $N=20$
was successfully probed as an enhancement of the $\sigma_R$ value.
A systematic measurement of the cross sections
along the Zr isotopic chain sheds light on the emergence
of the quadrupole deformation. We will discuss later its capability
as a quantitative probe of the quadrupole deformation of the Zr region.

In this paper, we generate the density distributions of
neutron-rich even-even Zr isotopes by
the Skyrme-Hartree-Fock-Bogoliubov (HFB) model using
the HFB solver, HFBTHO~\cite{Stoitsov13}. 
The SkM$^\ast$ interaction is employed~\cite{SkMs}
as it is known to reasonably describe nuclear deformations. 
A standard mixed-type pairing interaction is used~\cite{Sandulescu05}. 
Note that the HFBTHO code assumes an axial symmetry in the nuclear density. 
To access the effects of this constraint and the pairing interaction,
we compute the Skyrme-HF plus
Bardeen-Cooper-Schrieffer(BCS)-type pairing (HF+BCS)
in three-dimensional coordinate space. 
The BCS pairing strength is adjusted to reproduce
the pairing rotational energy around $^{90}$Zr~\cite{Hinohara16}.
To see the effect of nuclear deformation, we also perform
the spherical-constrained HF calculation 
with the filling approximation~\cite{Beiner75}.

Once we obtain the intrinsic density distributions,
the density distributions in the laboratory frame
are obtained by taking an average over angles
as prescribed in Ref.~\cite{Horiuchi12}.
This density distribution is used as a direct input to
the calculation of $\sigma_R$.
Here we evaluate $\sigma_R$ on a carbon target
as it has the sensitivity of the nuclear density profiles
near the nuclear surface~\cite{Horiuchi14}.
The Glauber theory~\cite{Glauber} with the nucleon-target formalism~\cite{NTG} is used to evaluate the optical phase-shift function.
The inputs to the theory are the density distributions of the projectile
and target nuclei and the profile function.
A standard parameter set of the profile function is tabulated
in Ref.~\cite{Ibrahim08} and has been well tested in high-energy nucleus-nucleus collisions including unstable nuclei~\cite{Horiuchi06,Horiuchi07,Ibrahim09,Horiuchi10,Horiuchi12,Horiuchi15,Horiuchi16,Horiuchi17,Nagahisa18}.
Thus, $\sigma_R$ properly reflects
the information on the density profile of the projectile nucleus.

\begin{figure}[ht]
\begin{center}
  \epsfig{file=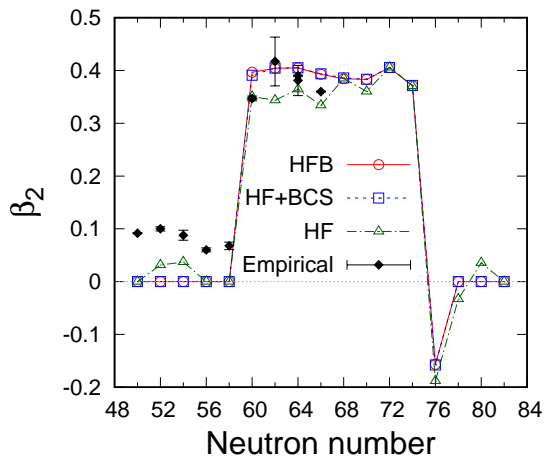, scale=1.2}
  \caption{Quadrupole deformation parameters $\beta_2$
    of even-even Zr isotopes as a function of neutron number
    calculated with the HFB, HF+BCS, and HF models.
    Experimental $|\beta_2|$ values 
    are taken from Refs.~\cite{NuDat3, Browne15}.
}
    \label{beta2.fig}
  \end{center}
\end{figure}

Figure~\ref{beta2.fig} displays calculated quadrupole deformation parameters
$\beta_2$ of even-even Zr isotopes with $N=50$--$82$ together
with experimentally extracted $|\beta_2|$ values~\cite{NuDat3, Browne15}. 
A negative $\beta_2$ value means oblate deformation.
The isotopes with $N=50$--58 are spherical
and show large and sudden enhancement of the $\beta_2$ value
at $N=60$. This large deformation $\beta_2 \approx 0.4$
is kept up to $N=74$. At $N=76$,
the deformation of the ground state is suddenly changed
to an oblate shape, $\beta_2 \approx -0.2$,
and the isotopes at $N=78$--$82$ become spherical again.
The emergence of nuclear deformation will be discussed later in detail.

Though there are small quantitative differences,
all the theoretical models, HFB, HF+BCS, and HF, 
show similar isotope dependence and
reproduce the trend of the experimental data,
especially, the sudden onset of the deformation at $N=60$.
Note that this large prolate deformation is also supported by
the large quadrupole deformation $\beta_2=0.4283(317)$ deduced from
the $E2$ moment measurement for $^{101}$Zr ($N=61$)~\cite{Campbell02}.

We note that the HFB and HF+BCS results are almost identical,
indicating that the nuclear deformation and pairing correlations
in Zr isotopes are well described in the HF+BCS approach as well.
The HF results, which do not include the pairing correlations,
mostly follow the HFB and HF+BCS ones but some deviations
are found. This is because the HF calculations exhibit triaxial deformation:
$\gamma=14^\circ, 14^\circ$, and 20$^\circ$ for these nuclei
with $N=62, 64$, and 66, respectively.

We confirm that no triaxial deformation is obtained
in the HF+BCS calculations. The resulting ground-state wave function
is found to be an axially symmetric, prolate or oblate shape,
i.e., the $\gamma$ value is 0$^\circ$ or $60^{\circ}$.
The pairing interaction plays a role in suppressing
the triaxiality of Zr isotopes in this mass region.

\begin{figure}[ht]
\begin{center}
  \epsfig{file=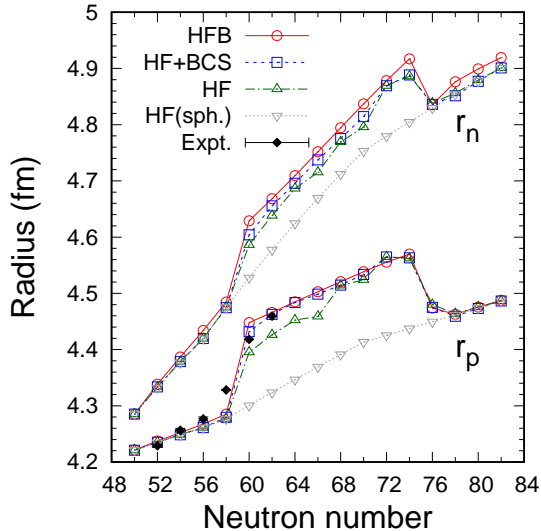, scale=1.2}
  \caption{Rms point-proton ($r_p$) and neutron ($r_n$) radii
    of even-even Zr isotopes as a function of neutron number
    calculated with the HFB, HF+BCS, HF,
    and spherical-constrained HF models.
    The experimental point-proton radii are taken from
    Ref.~\cite{Angeli13}.}
    \label{radii.fig}
  \end{center}
\end{figure}

It is well known that the isotope dependence of nuclear radii follows
that of the nuclear quadrupole deformation in light to medium mass nuclei~\cite{Minomo11,Sumi12,Minomo12,Horiuchi12,Takechi14,Watanabe14,Horiuchi15,Horiuchi21,Choudhary21, Horiuchi22}. 
Figure~\ref{radii.fig} plots the root-mean-square (rms) point-proton ($r_p$)
and neutron ($r_n$) radii of even-even
Zr isotopes calculated by the HFB, HF+BCS, HF,
and spherical-constrained HF models.
The experimental point-proton radii~\cite{Angeli13}
are also plotted for comparison.
The results with the HFB and HF+BCS models
nicely reproduce the experimental values,
while the ones with the spherical-constrained
HF calculation do not explain
the experimental proton radii at $N=60$ and 62.
The nuclear deformation is essential to account for the proton radii.
At a closer look, the HF results show
small deviations as their quadrupole deformation
are not large enough as shown in Fig.~\ref{beta2.fig} 
due to the lack of the pairing correlations.

The reproducibility in $\beta_2$ and $r_p$ clearly
shows the validity of the present theoretical models. 
We note that the point-neutron radii also behave like the proton ones.
This coincident trend of the proton and neutron radii
suggests that both protons and neutrons
associate to develop nuclear deformation.
  
\begin{figure}[ht]
\begin{center}
  \epsfig{file=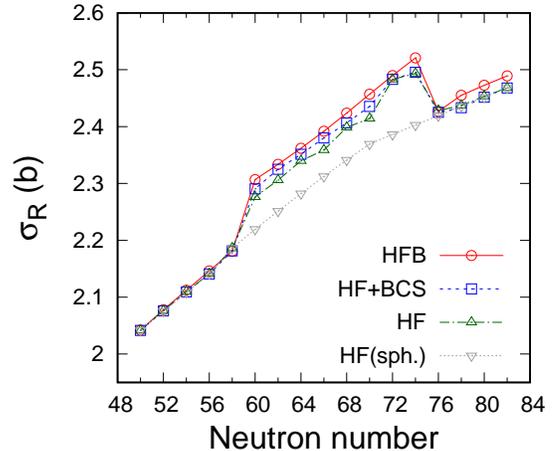, scale=1.2}                    
  \caption{Total reaction cross sections on a carbon target
    of even-even Zr isotopes at 300 MeV/nucleon as
    a function of neutron number
    calculated with the HFB, HF+BCS, HF, and spherical-constrained HF models.}
    \label{rcs.fig}
  \end{center}
\end{figure}

\begin{figure}[ht]
\begin{center}
  \epsfig{file=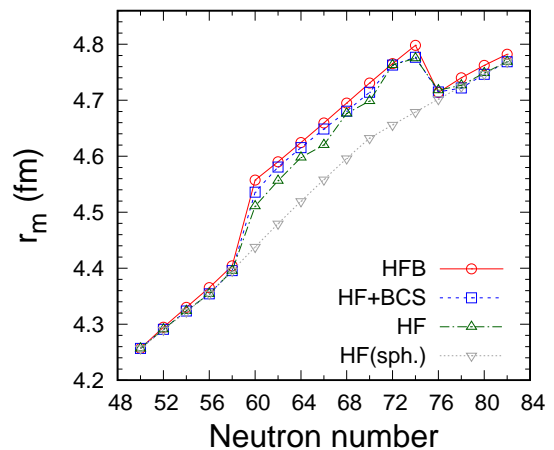, scale=1.2}  
  \caption{Same as Fig.~\ref{radii.fig} but for matter radii ($r_m$).}
    \label{radiim.fig}
  \end{center}
\end{figure}

As mentioned before, these density profiles can be used
as the inputs to the Glauber theory, allowing us to
relate the rms matter radius $(r_m)$
to a high-energy reaction observable, $\sigma_R$.
Figure~\ref{rcs.fig} plots $\sigma_R$ on a carbon target at 300 MeV/nucleon,
which is a typical condition of
the recent systematic measurements~\cite{Takechi12,Takechi14,Tanaka20}.
The trend of $\sigma_R$ reflects the behavior of the matter radii
displayed in Fig.~\ref{radiim.fig}:
a sudden enhancement at $N=60$ and reduction at $N=76$ compared to
the spherical one, which only shows monotonic increases of the matter radii.
In $\sigma_R$, this deformation effect, i.e.,
the difference from the spherical-constrained HF calculation, is approximately 100~mb,
which is enormously large recalling that the cross-section increase
in Ne and Mg isotopes are typically a few tens of mb,
at most $\approx 50$ mb~\cite{Horiuchi12}. 
Since this enhancement corresponding to the relative change of $4$\% ($\approx 100\mbox{ }\mathrm{mb}/2500\mbox{ }\mathrm{mb}$) is large enough to distinguish experimentally under the typical experimental condition~\cite{Takechi12,Takechi14,Tanaka20}, it is anticipated that the region beyond $N=70$, which is the subshell closure of $sdg$ orbits, which will be focused on later, can be accessed through the $\sigma_R$ measurement with sufficient experimental precision using the high-intensity radioactive ion beams of 0.1--1 counts per second
in which the recent sophisticated facility can be supplied~\cite{Ohnishi10,Shimizu18}.
On the other hand, it seems difficult to reach the $N=70$ isotope for the moment by using the $\beta$-$\gamma$ delay coincidence technique, 
that was adopted for the $B(E2)$ measurement of $^{106}$Zr~\cite{Browne15}
because a parent Y isotope production rapidly decreases
  as the neutron number increases.

The mechanism of the large cross-section enhancement due to
large quadrupole deformation
can be explained by the occupation of the intruder orbits.
In such a largely deformed region, the asymptotic quantum number
$[nn_z\Lambda]\Omega$ is useful to classify the single-particle
orbits~\cite{Nilsson55}.
In $60\leq N\lesssim 70$, 
the [550]1/2 orbit, which comes from the spherical $0h_{11/2}$ orbit,
plays a role to induce the large quadrupole deformation.
When the spherical $sdg$ orbits are fully occupied at $N \approx 70$,
the nuclear system will be a spherical shape
if the shell gap between the spherical $sdg$ and $0h_{11/2}$ orbits is sizable.
However, the large nuclear deformation is kept up to $N=74$.
Another mechanism is needed for explaining this enlargement of the deformation region.
We remark that the enlargement of the deformation region at $N \approx 72$
in Zr isotopes are seen in several previous papers~\cite{Geng03,Miyahara18} 
but its mechanism was not discussed.

\begin{figure}[htb]
  \centering
\epsfig{file=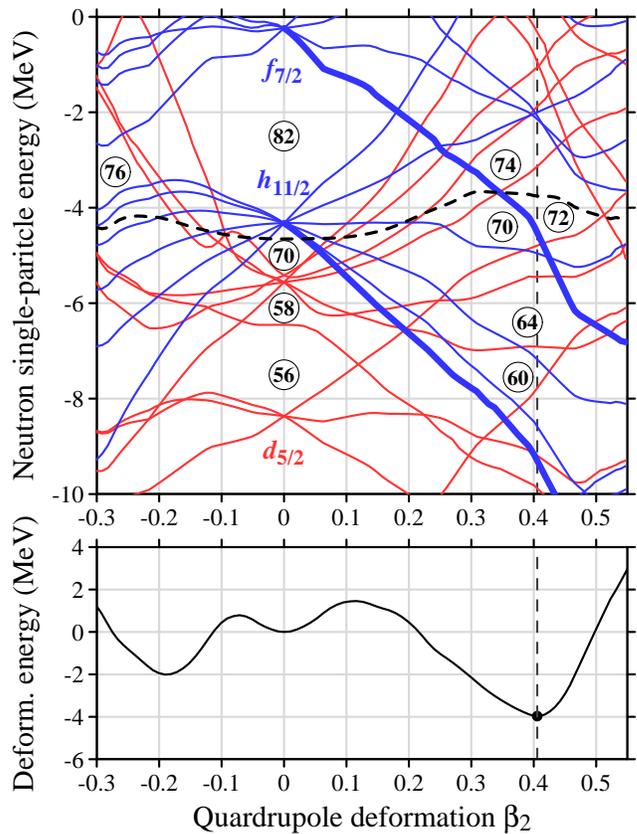,scale=0.4}
\caption{Neutron single-particle 
  and deformation energies of $^{112}$Zr as a function
  of quadrupole deformation parameter $\beta_2$.
  A thick dashed curve denotes the neutron chemical potential.
  Thick solid curves indicate the $[550]1/2$ and $[530]1/2$ orbits, which comes from the spherical $0h_{11/2}$ and $1f_{7/2}$ orbits $(\beta_2=0)$, respectively.
  A vertical thin dashed line indicates the $\beta_2$ value
  at the energy minimum which corresponds
  to the ground state of $^{112}$Zr for a guide to the eye.
}
\label{config.fig}
\end{figure}

To clarify the reason for the enlarged deformation region at $N=72$--74,
we plot in Fig.~\ref{config.fig} the neutron single-particle energies
of $^{112}$Zr as a function of the quadrupole deformation parameter.
The deformation energy, which is measured from that with the spherical shape ($\beta_2=0$),
is also plotted as a guide.
$^{112}$Zr has the energy minimum at $\beta_2=0.4$.
Since the nuclear deformation is so large,
the elongated, next intruder orbit with
$[530]1/2$ coming from the spherical $1f_{7/2}$ orbit
is occupied and further enhances the nuclear radius.
This ``second'' intruder orbit acts to form the shell gaps
at $N=72$ and $74$, and hence such a large deformation is stabilized. 
On the other hand, this single-particle energy diagram
indicates that the shell gap at $N=76$ does not appear
in prolate deformation but in oblate deformation around
$\beta_2 \approx -0.2$.
This situation may signature that the sudden change
from prolate to oblate deformation at $N=76$.

We remark that this sort of the enlargement of deformation regions
can also be seen in Mg isotopes.
In the neutron-rich Mg isotopes, the island of inversion starts
from $^{31}$Mg with the occupation of the [330]1/2
orbit originating from $0f_{7/2}$.
We see that the $^{40}$Mg nucleus is still largely deformed $\beta_2=0.3$.
Compared to that in Ne isotopes ($18<N\leq 24$),
the deformation region of Mg isotopes is enlarged to $N=28$
due to the occupation of the ``second'' intruder orbit with [310]1/2
from $1p_{3/2}$. On the other hand, for Ti, Cr, and Fe isotopes,
the magicity of the spherical $0g_{9/2}$ orbit ($N=50$) is strong and 
the large deformation is suppressed before
the next intruder orbit from $1d_{5/2}$ contributes.
This can be attributed to the large level spacing of
the first and second intruder orbits at the spherical limit,
which respectively comes from $0g_{9/2}$ and $1d_{5/2}$.
We expect that the same phenomenon occurs
at heavier neutron-rich deformed regions
such as the northeast area of $^{132}$Sn, 
in which the next intruder from $1g_{9/2}$ orbit
would contribute at $N \approx 110$.

\begin{figure}[ht]
\begin{center}
  \epsfig{file=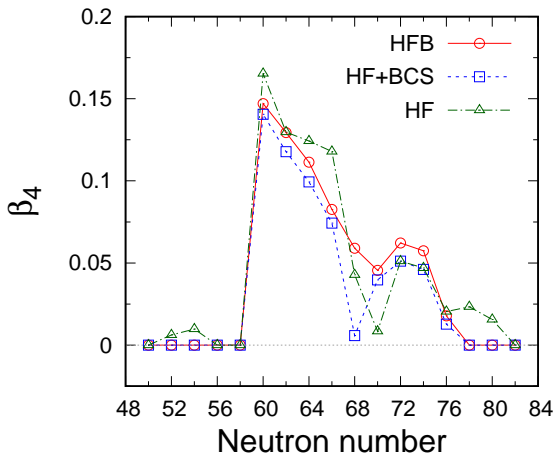, scale=1.2}                    
  \caption{Hexadecapole deformation parameters $\beta_4$ of even-even Zr isotopes as a function of neutron number calculated with the HFB, HF+BCS, and HF models.}
    \label{beta4.fig}
  \end{center}
\end{figure}

We discuss the effect of the occupation of the second intruder orbits
on the density profile.
Here we examine a higher order of the nuclear shape, the hexadecapole deformation parameter, 
which can be an indicator of the occupation of
an elongated orbit~\cite{Horiuchi22}. 
Figure~\ref{beta4.fig} shows the hexadecapole deformation parameters
$\beta_4$ as a function of the neutron number. 
As expected, the $\beta_4$ value is suddenly enhanced at $N=60$ due to the occupation of the [550]1/2 orbit, 
which gives the maximum hexadecapole moment among the orbits with the principal quantum number $n=5$~\cite{BM}.
As the neutron number increases, the $\beta_4$ value is reduced
by the occupation of less elongated orbits. 
We see a kink behavior at $N\approx 70$. The sizable effect on the $\beta_4$ value is found by the occupation of the elongated intruder orbit with $[530]1/2$
though the degree is smaller than that with [550]1/2. 
It is desirable to know the isotope dependence of the $\beta_4$ values
in Zr isotopes experimentally.
 However, as detailed in Ref.~\cite{Horiuchi22}, a direct
  determination of $\beta_4$ is not easy by the total reaction cross section
  measurement based on a simple deformed density model. An elaborated
  analysis is needed to establish it and this is the next scope of our study.

\begin{figure}[ht]
\begin{center}
  \epsfig{file=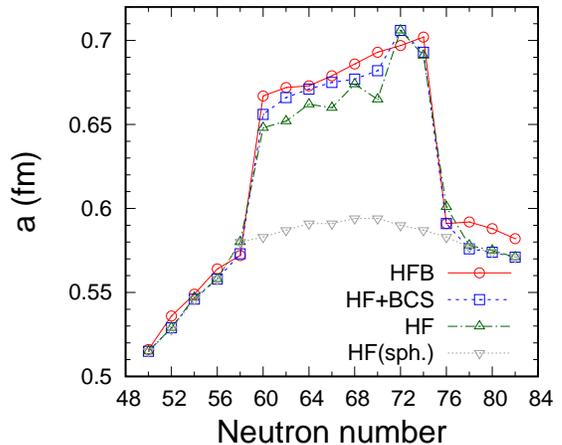, scale=1.2}
  \caption{Nuclear diffuseness of even-even Zr isotopes
    as a function of neutron number calculated
    with the HFB, HF+BCS, HF, and spherical-constrained HF models. 
    See text for details.}
    \label{diff-Zr.fig}
  \end{center}
\end{figure}

  Additionally, this characteristic behavior of $\beta_{2}$ and
  $\beta_{4}$ is certainly
  reflected in the density distribution near the nuclear surface,
  and its evaluation is meaningful. To quantify the surface density profile,
  we compute the nuclear diffuseness, which is practically obtained
  by minimizing the difference between the calculated density distribution
  and the two-parameter Fermi function with radius ($R$)
  and diffuseness ($a$) parameters. These parameters can be extracted
  almost uniquely by measuring the proton-nucleus elastic scattering
  differential cross section at the first diffraction
  peak~\cite{Hatakeyama18}. The thus-obtained $a$ value can be a good measure
  of the density profile near the nuclear surface and includes a variety of nuclear structure information
such as nuclear deformation~\cite{Choudhary21,Horiuchi22},
clustering \cite{Horiuchi22c, Horiuchi23},
bubble structure~\cite{Choudhary20},
and core swelling phenomena~\cite{Tanaka20,Horiuchi20,Horiuchi22b}.
Figure~\ref{diff-Zr.fig} displays the diffuseness parameters of Zr isotopes extracted from the matter density distributions.
If the nuclear system exhibits some deformation, the diffuseness value
is enhanced in most cases~\cite{Choudhary21,Horiuchi21,Horiuchi22}
due to the occupation of the nodal momentum orbits
near the Fermi level~\cite{Horiuchi21b}.
In the spherical region at $50\leq N \leq 60$,
the $a$ values fall within the range of the standard value
about 0.5--0.6 fm~\cite{BM,Horiuchi22},
while it becomes an extremely large value $a \approx 0.7$ fm
when the system exhibits large deformation.
 The behavior at
  $N\approx 70$--74 is amplified in the $a$ parameters,
  and therefore the nuclear diffuseness is also sensitive to the occupation
  of the second intruder orbit and is
  another possible way to probe the characteristic behavior of
  nuclear deformation.

Extending these analyses made for Zr isotopes,
we also evaluate the quadrupole deformations of neighboring isotopes
for a more general understanding of this phenomenon.
We find the enlargement of the deformation region in Sr isotopes,
which shows similar deformation properties of Zr isotopes. 
On the other hand, Kr, Mo, and Ru isotopes exhibit some triaxial deformations
with less quadrupole deformation compared to the Zr and Sr isotopes.
The enlargement of the quadrupole deformation region
by the second intruder orbits does not occur in those isotopes.

In summary, we have investigated the nuclear deformation and density profiles of neutron-rich even-even Zr isotopes. The density distributions are generated by the Skyrme-Hartree-Fock-Bogoliubov theory, which can treat both nuclear deformation and paring correlations appropriately. The calculated results are consistent with the quadrupole deformation parameters and point-proton radii deduced from experiments.
The anomalously large enhancement of the total reaction cross sections
is predicted at $N=60$ owing to the occupation of the highly elongated intruder
orbit originating from spherical $0h_{11/2}$ orbit.
As the neutron number increases, the isotopes keep their large deformation up to $N=74$. The deformation region is enlarged by the next intruder orbits originating from $1f_{7/2}$ orbits. The signature of the occupation of these intruder orbits can be detected by investigating the nuclear hexadecapole deformation and the density profiles near the nuclear surface. The latter can be observed by using the nucleus-proton elastic scattering. It is highly desirable to systematically measure the total reaction and elastic scattering cross sections for understanding the mechanism of the enlargement of deformation regions and
the radius enhancement in heavy mass nuclei $A\gtrsim 110$.

%\acknowledgments
This work was in part supported by JSPS KAKENHI Grants Nos.\ 18K03635
and 22H01214. We acknowledge the Collaborative Research Program 2022, 
Information Initiative Center, Hokkaido University.

\end{document}